\documentclass[aps,prl,twocolumn,superscriptaddress,english]{revtex4-2}
\usepackage[utf8]{inputenc}
\usepackage[english]{babel}
\usepackage{hyphenat}
\usepackage{amsmath}
\usepackage{amssymb}
\usepackage{graphicx}
\usepackage{xcolor}
\usepackage{siunitx}
\usepackage{mathtools}
\usepackage{lineno}

\usepackage{hyperref}

\newcommand{\Tc}{$T_\text{c}$}

\newcommand{\pkin}{$p_\text{kin}$}
\newcommand{\pdyn}{$p_\text{dyn}$}
\newcommand{\BSH}{BaSiH$_8$}
\newcommand{\SSH}{SrSiH$_8$}
\newcommand{\LBH}{LaBH$_8$}
\newcommand{\XYH}{$XY$H$_8$}
\newcommand{\fmm}{$Fm\bar{3}m$}
\newcommand{\qv}{\mathbf{q}}
\newcommand{\kv}{\mathbf{k}}
\newcommand{\SSCHA}{\mbox{SSCHA}}

\makeatletter
\def\@fnsymbol#1{\ensuremath{\ifcase#1\or \dagger\or *\or \ddagger\or
   \mathsection\or \mathparagraph\or \|\or **\or \dagger\dagger
   \or \ddagger\ddagger \else\@ctrerr\fi}}
\makeatother

\DeclareSIUnit\angstrom{\protect \text {\AA}}

\begin{document}

\author{Roman Lucrezi}
\affiliation{Institute of Theoretical and Computational Physics, Graz University of Technology, NAWI Graz, 8010 Graz, Austria}
\author{Eva Kogler}
\affiliation{Institute of Theoretical and Computational Physics, Graz University of Technology, NAWI Graz, 8010 Graz, Austria}
\author{Simone Di Cataldo}
\affiliation{Institute of Theoretical and Computational Physics, Graz University of Technology, NAWI Graz, 8010 Graz, Austria}
\affiliation{Dipartimento di Fisica, Sapienza Universit\`a di Roma, 00185 Rome, Italy} 
\author{Markus Aichhorn}
\affiliation{Institute of Theoretical and Computational Physics, Graz University of Technology, NAWI Graz, 8010 Graz, Austria}
\author{Lilia Boeri} 
\affiliation{Dipartimento di Fisica, Sapienza Universit\`a di Roma, 00185 Rome, Italy}
\affiliation{Enrico Fermi Research Center, Via Panisperna 89 A, 00184 Rome, Italy}
\author{Christoph Heil} \thanks{Corresponding author} \email{christoph.heil@tugraz.at}
\affiliation{Institute of Theoretical and Computational Physics, Graz University of Technology, NAWI Graz, 8010 Graz, Austria}

\title{Quantum lattice dynamics and their importance in ternary superhydride clathrates}

\date{\today}
\begin{abstract}
\textbf{Abstract:}
The quantum nature of the hydrogen lattice in superconducting hydrides can have crucial effects on the material's properties. Taking a detailed look at the dynamic stability of the recently predicted \BSH\ phase, we find that the inclusion of anharmonic quantum ionic effects leads to an increase in the critical dynamical pressure to \SI{20}{GPa} as compared to \SI{5}{GPa} within the harmonic approximation. We identify the change in the crystal structure due to quantum ionic effects to be the main driving force for this increase and demonstrate that this can already be understood at the harmonic level by considering zero-point energy corrections to the total electronic energy. In fact, the previously determined critical pressure of kinetic stability \mbox{\pkin\ = \SI{30}{GPa}} still poses a stricter bound for the synthesizability of \BSH\ and similar hydride materials than the dynamical stability and therefore constitutes a more rigorous and accurate estimate for the experimental realizability of these structures.
\end{abstract}

\maketitle

\section{Introduction}
The discovery of high-temperature superconductivity in H$_3$S at extreme pressures~\cite{drozdov2015conventional} stimulated an intense hunt for novel hydride compounds with even higher critical temperatures (\Tc), spearheaded by computational material discovery~\cite{Boeri_2019,Lilia_2022,FLORESLIVAS20201,Zurek_Hilleke2022,Zurek_chemPrecomp2022}. One prominent example is LaH$_{10}$, which has been shown to superconduct up to temperatures of \SI{265}{K} at pressures of $\sim$\SI{190}{GPa}~\cite{drozdov2019superconductivity,somayazulu2019evidence}.
While it is very tempting to continue searching for materials with record-breaking \Tc's~\cite{peng2017hydrogen, grockowiak2020hot, DiCataldo_2022}, lowering the required stabilization pressures is even more important in view of technological applications~\cite{pickard2020superconducting,lv2020theory,di2020phase,DiCataldo_2021,shipley2021high,zhang2021design,PhysRevB.104.134501,Lucrezi2022,DiCataldo_2022,Sun_2022}.

In a recent paper~\cite{DiCataldo_2021} some of us proposed a strategy to bring the stabilization pressures of high-\Tc\ hydrides closer to ambient pressure, based on the concept of an optimized \textit{chemical precompression}. In fact, we identified \LBH, a hydride superconductor with a \Tc\ $>$ \SI{100}{K}, dynamically stable at an unprecedentedly low pressure.
In a follow-up work, we showed that other hydride superconductors with the same \fmm\ \XYH\ structural template can be identified with even lower critical pressures of stability, such as \SSH\ and \BSH~\cite{Lucrezi2022}. The latter is particularly interesting, as it remains dynamically stable down to \SI{3}{GPa}. We note in passing that these estimates of dynamical stability were based on anharmonic frozen-phonon calculations. 

All of the mentioned \fmm\ \XYH\ compounds are thermodynamically stable only at pressures above $\sim$\SI{100}{GPa}, and metastable below. A conceivable route to synthesize these materials would hence be to obtain them at high pressures where they are thermodynamically stable, and quench them to lower pressures. Such a procedure has recently been successfully employed to realize the \fmm\ LaBeH$_8$ at around \SI{110}{GPa} with subsequent quenching down to a pressure of $\sim$\SI{80}{GPa}~\cite{Song_2023_stoichiometric}.

The standard criterion employed in literature to estimate how far a metastable phase can be quenched down in pressure is \textit{dynamical} (phonon) stability. However, dynamical stability indicates only that a structure is in a local minimum of the potential energy surface. To estimate its actual lifetime (\textit{kinetic} stability) one needs also to estimate the height of the barriers that separate the current minimum from other minima. In a previous work~\cite{Lucrezi2022}, some of us introduced a rigorous method to assess the kinetic stability pressure \pkin\ by explicitly calculating the energy barrier protecting the metastable \fmm\ structure from decomposition as a function of pressure, using the variable-cell nudged elastic band method~\cite{QIAN20132111}. For \BSH, for example, we found a \pkin\ of $\sim$\SI{30}{GPa}, significantly higher than the dynamical value \pdyn\ = \SI{3}{GPa}. 

It was argued in a recent work~\cite{Errea_LaBH8} that quantum lattice effects treated within the stochastic self-consistent harmonic approximation (\SSCHA) drastically increase the dynamical stabilization pressure \pdyn\ for \LBH\ and it was further suggested that a similar increase in \pdyn\ should be expected for other \fmm\ \XYH\ hydrides. 

To investigate this, we apply the \SSCHA\ formalism to \BSH, which, so far, has the lowest \pdyn\ among all \fmm\ \XYH\ hydrides. 
In \SSCHA, a major bottleneck is represented by the need to use large supercells and large numbers of randomly displaced structures if one wants to fully converge the calculation. We overcome this problem by employing machine-learned moment tensor potentials (MTP)~\cite{Shapeev2015-MTP,GUBAEV2019148} that allow us to obtain total energies, forces, and stresses with density-functional-theory (DFT) accuracy but at a fraction of the computational cost~\cite{ranalli_SSCHA,D1NR08359G,MLIP_eval}. To our knowledge, this work represents the first combination of MTPs with the \SSCHA\ method.
In addition, we introduce a method to discern the contribution of quantum ionic (QI) effects from those of anharmonic (anh) and phonon-phonon (ph-ph) effects.

We find that \pdyn\ increases from \SI{3}{GPa} to about \SI{20}{GPa} within the \SSCHA\ and that this rise can almost entirely be attributed to QI effects, with actual anharmonic and ph-ph effects playing only a subordinate role. 
In fact, the same crystal structure that minimizes the free energy within \SSCHA\ can already be obtained at the harmonic level using DFT by including zero-point energies (ZPE).

Particularly, we demonstrate that even after including QI, anharmonic, and ph-ph effects within the framework of \SSCHA, the actual limit of stability is still set by \pkin\ ($\sim$\SI{30}{GPa}), as stated in our previous work~\cite{Lucrezi2022}.

\section{Results}
\subsection{Ab-initio machine-learned interatomic potentials}
In the self-consistent harmonic approximation, the system of fully anharmonic and interacting lattice vibrations is mapped onto an
auxiliary harmonic system and the free energy $\mathcal{F}$ of the full system is approximated by the minimum of the free energy of the auxiliary harmonic system~\cite{Born1951,SCHA_Hooton,PhysRevLett.17.89,PhysRevB.1.572}. In the \SSCHA, this minimization is performed stochastically via Monte-Carlo summation and importance sampling over several consecutive ensembles (populations) of a large number of individuals. Each individual here corresponds to a supercell structure with displaced atomic positions, where the supercell size determines the density of phonon wave vectors in the Brillouin zone~\cite{PhysRevB.89.064302}. More details are provided in the Method section, in Supplementary Method~2, and in Refs.~\cite{PhysRevLett.111.177002,PhysRevB.96.014111,PhysRevB.97.014306,PhysRevB.98.024106,PhysRevLett.122.075901,errea2020quantum,Monacelli_2021}. 

In practice, to calculate accurate phonon frequencies within \SSCHA, in particular for slow-converging soft modes, one needs to consider population sizes of several ten or hundred thousands individuals. In addition, one also needs to converge the supercell size. 

Doing this fully at a DFT level is computationally prohibitive, which is why we made use of MTPs in this work. For every pressure, MTPs were trained on DFT results of 50 structures randomly chosen out of the \SSCHA\ random-displacement individuals in 2$\times$2$\times$2 supercells. We then validated the trained MTPs for all other individuals by comparing the total energies, forces and stress components. This validation is shown in Fig.~\ref{fig:MLIP_valid}, demonstrating the exceptional accuracy of the used MTPs (see Supplementary Fig.~2 for other pressures, as well as forces and stresses). As can be appreciated in this figure, the root-mean-squared error (RMSE) is below \SI{1}{meV/atom}, i.e., at the same level as the error in DFT. The inset also shows that the potential energy surface of the slow-converging, $T_{2g}$ mode at $\Gamma$ is reproduced very nicely with the MTPs.

As a final validation, we compare the \SSCHA\ phonon dispersions obtained using only DFT with those employing only MTPs (see Supplementary Fig.~3) and find very good agreement, with only minor differences in the $T_{2g}$ mode at $\Gamma$ and the $E_g$ at $X$. To fully converge these modes within \SI{1}{meV}, we increased the populations sizes within MTP-\SSCHA\ up to \SI{100000}{} individuals compared to \SI{10000}{} for the DFT-\SSCHA\ calculations.

\begin{figure}[t]
\centering
	\includegraphics[width=\columnwidth]{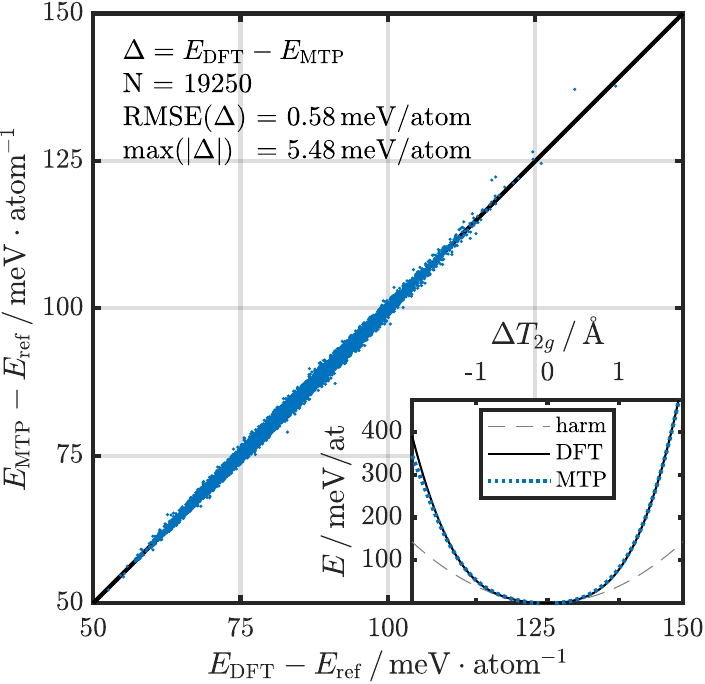}
	\caption{\textbf{MTP validation:} MTP total energy ($E_\mathrm{MTP}$) versus DFT total energy ($E_\mathrm{DFT}$) for all $N$ = \SI{19250}{} individuals of a \SSCHA\ calculation for a lattice constant $a=\SI{6.242}{\angstrom}$ (blue scatter plot). $E_\mathrm{ref}$ is the DFT total energy of the high-symmetry structure with undisplaced H positions. The diagonal black line serves as guide to the eye. The inset shows the full frozen-phonon potential of the lowest $T_{2g}$ mode at $\Gamma$ obtained with DFT (solid black line) and MTP (dotted blue line), as well as the harmonic potential (dashed grey line). The root-mean-squared error (RMSE) and the maximum value of the prediction error $\Delta = E_\text{DFT}-E_\text{MTP}$ for the presented data set is shown in the upper left corner.}
	\label{fig:MLIP_valid}
\end{figure}

The use of MTPs does not only substantially speed up the calculations (we found MTPs to be a factor of about \SI{20000}{} faster than DFT in the case of 2$\times$2$\times$2 supercells of \BSH), but also gives access to larger supercells. In this work, we performed additional \SSCHA\ calculations using MTPs on $n\times n\times n$ supercells with $n=1,2,3,4$ at all studied pressures. The convergence of the free energy, the structural parameters, and the phonon dispersions with respect to the supercell size is provided in Supplementary Figs.~4-6. An overview of all performed \SSCHA\ runs is given in Supplementary Tab.~1.
Unless stated otherwise, all \SSCHA\ results presented in the following have been obtained with MTPs for \SI{100000}{} individuals in 4$\times$4$\times$4 supercells at \SI{0}{K}.

\subsection{Structural parameters and electronic dispersion}
The \fmm\ phase of \BSH\ has a face-centered cubic unit cell with 10 atoms in the primitive cell, where Ba and Si occupy Wyckoff $4a/b$ sites and the H occupy $32f$ sites. The eight H atoms form rhombicuboctahedral cages around the Ba atoms and cubic cages around the Si atoms. The structure has only two free parameters, namely the lattice constant $a$ and the Wyckoff coordinate of the $32f$ sites $x$, defining the H-H distance $d_{\text{H-H}}=2a\cdot x$ (side length of the cubic cage) and the H-Si distance $d_{\text{H-Si}}=a\sqrt{3}\cdot x$ (half the space diagonal of the cubic cage).  

Relaxing the structure within DFT to target pressures of 10, 20, 25, and \SI{30}{GPa}, we obtained lattice constants between \SI{6.5}{\angstrom} and \SI{6.2}{\angstrom}, and H-Si distances of about \SI{1.6}{\angstrom}, as shown in Table~\ref{tab:sructpar}. An extensive list of the structural parameters from ambient pressure up to \SI{100}{GPa}, as well as the fit to the Birch-Murnaghan equation of state can be found in Supplementary Fig.~1 and Supplementary Note~1.

Starting from the atomic positions obtained in DFT and the harmonic dynamical matrices obtained in density-functional perturbation theory (DFPT) calculations at each pressure, we performed constant-volume \SSCHA\ relaxation calculations.
The corresponding parameters, indicated by $\tilde{x}, \tilde{d}$, and $\tilde{p}$, are reported in Table~\ref{tab:sructpar}. 

\begin{table}[t]
	\caption{\textbf{Structural parameters:} Lattice constant $a$, Wyckoff parameter $x$, H-Si distance $d_{\text{H-Si}}$, and pressure $p$ after the relaxation with respect to the DFT total energy and after the constant-volume relaxation within \SSCHA\ ($\tilde{x}, \tilde{d}$, and $\tilde{p}$).}
	\label{tab:sructpar}
\begin{tabular}{c|ccc|ccc}
 $a\,/\,\text{\AA}$ & $x$  & $d_{\text{H-Si}}\,/\,\text{\AA}$  & $p\,/\,\text{GPa}$  & $\tilde{x}$  & $\tilde{d}_{\text{H-Si}}\,/\,\text{\AA}$ & $\tilde{p}\,/\,\text{GPa}$\\
 \toprule
6.541 & 0.1434 & 1.625 & 10  & 0.1459 & 1.653 & 12.2\\
6.323 & 0.1471 & 1.611 & 20  & 0.1498 & 1.640 & 22.6 \\ 
6.242 & 0.1483 & 1.603 & 25  & 0.1510 & 1.633 & 27.9 \\
6.171 & 0.1494 & 1.597 & 30  & 0.1521 & 1.626 & 33.1
\end{tabular}
\end{table}

\begin{figure}[h]
	\includegraphics[width=\columnwidth]{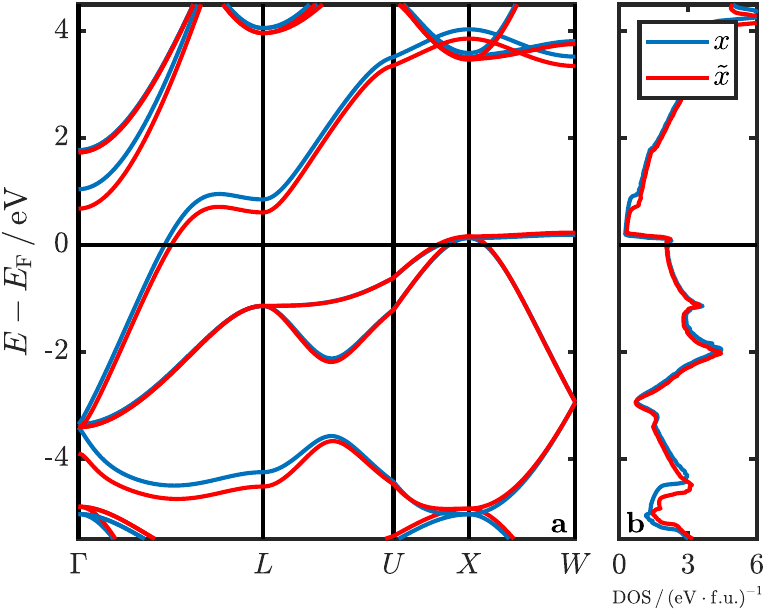}
	\caption{\textbf{Difference in electronic properties:} \textbf{a} electronic bands, and \textbf{b} density of states for the structure with H positions defined by $x$ (DFT minimum, blue line) and $\tilde{x}$ (\SSCHA\ minimum, red line) for $a=\SI{6.242}{\angstrom}$. The legend in \textbf{b} also applies to \textbf{a}.}
	\label{fig:el}
\end{figure}

We observe an elongation of $d_{\text{H-H/Si}}$ of about \SI{30}{\milli\angstrom} (2\%) for all pressures and an increase in pressure of about 2 to \SI{3}{GPa}, i.e. $\sim20\%$ at \SI{10}{GPa} and $\sim10\%$ at \SI{30}{GPa}.
The change in atomic positions introduces only small changes in the electronic structure, as demonstrated in Fig.~\ref{fig:el}, where we compare the electronic bands and densities of states (DOS) for $x$ and $\tilde{x}$.  
The largest differences are found above and below the Fermi energy, whereas electronic bands and DOS at the Fermi energy, and hence the Fermi surface, remain essentially unchanged.

\subsection{Phonon dispersions and lattice instability}
\begin{figure*}[ht!]
    \centering
    \includegraphics[width=\textwidth]{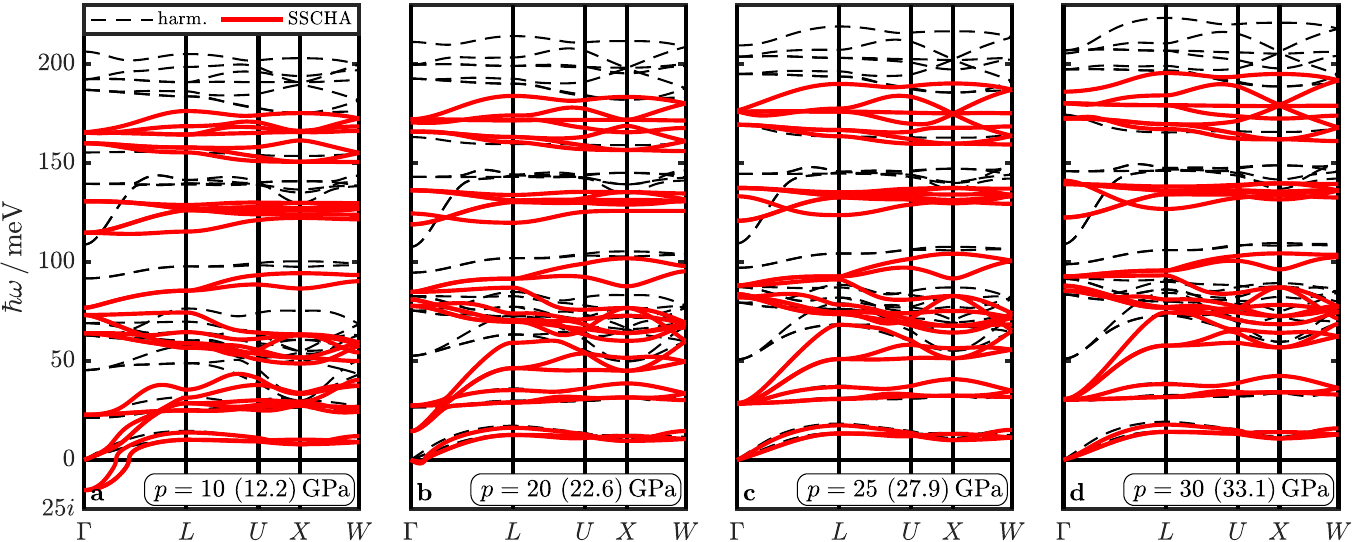}
    \caption{\textbf{Harmonic and \SSCHA\ phonon dispersions:} Phonon dispersions for various pressures along a high-symmetry path of the BZ. The dashed black lines correspond to harmonic calculations and the solid red lines to the \SSCHA\ results. The legend in \textbf{a} also applies to \textbf{b}, \textbf{c}, and \textbf{d}. The pressures indicated in \textbf{a}, \textbf{b}, \textbf{c}, and \textbf{d} correspond to the DFT (\SSCHA) pressures for relaxed atomic positions defined by $x$ ($\tilde{x}$). The related values for $x$, $\tilde{x}$, and the lattice constant $a$ are reported in Tab.~\ref{tab:sructpar}. The small imaginary dip near $\Gamma$ in \textbf{b} is due to interpolation, all phonon frequencies at wave vectors commensurate with the supercell are positive.}
    \label{fig:hessian_ph}
\end{figure*}

Moving on, we evaluate and compare the DFPT and \SSCHA\ phonon dispersions at all studied pressures, as shown in Fig.~\ref{fig:hessian_ph}. Similar to the results for \LBH\ reported by Belli et al.~\cite{Errea_LaBH8}, we find that the high-energy optical modes are strongly renormalized to lower frequencies. In particular, a significant softening occurs for the threefold degenerate $T_{2g}$ mode at $\Gamma$ (harmonic values around \SI{50}{meV} in Fig.~\ref{fig:hessian_ph}), which becomes imaginary and indicates a (dynamic) lattice instability for lattice constants \mbox{$a>\SI{6.323}{\angstrom}$}, corresponding to $p=\SI{20}{GPa}$ and $\tilde{p}=\SI{22.6}{\GPa}$. Thus, the inclusion of quantum lattice effects within the \SSCHA\ shifts the dynamical stability pressure from the anharmonic frozen-phonon value \mbox{\pdyn\ = \SI{3}{GPa}} to $\tilde{p}_\text{dyn} = \SI{20}{GPa}$ (see Supplementary Fig.~7). This $\sim$\SI{17}{GPa} difference is substantial, but considerably smaller than the $\sim$\SI{40}{GPa} shift reported for \LBH~\cite{Errea_LaBH8}. 

We want to note in passing that we also calculated the fourth order corrections to the phonon frequencies in \SSCHA\ (cf. Supplementary Method~2) and find, in contrast to the work on \LBH~\cite{Errea_LaBH8}, but in accordance with other works employing \SSCHA~\cite{Monacelli_2021,errea2020quantum,PhysRevB.96.014111,PhysRevLett.122.075901,PhysRevB.97.014306}, only minor differences to the results obtained only up to third order. The maximum phonon energy differences are on the order of about \SI{1}{meV} for all pressures. The phonon dispersions in the 2$\times$2$\times$2 supercells with second (auxiliary), third, and fourth order terms for all studied pressures are shown in Supplementary Fig.~8. 

\subsection{Different effects contributing to frequency shifts}
The observed changes in the phonon dispersions when employing \SSCHA\ and the resulting different dynamical stabilization pressures result from a combination of several effects that are not included at the level of standard DFT and DFPT. These are most importantly the vibrational contributions of the ions to the free energy, phonon anharmonicity, and ph-ph interactions. In the following, we will present an attempt to disentangle and determine the importance of each of these effects for \BSH.
Before doing so, however, we want to briefly touch upon terminologies around phonon anharmonicity. Anharmonicity gives rise to ph-ph interactions, but also to the phonon self-interaction of a single mode; effects that are sometimes collectively referred to as anharmonic effects. In a frozen-phonon approach, single-mode potentials are usually obtained without incorporating ph-ph interactions, and there anharmonicity then refers to any deviation from the idealized parabolic potential. For the sake of clarity, we mention both aspects explicitly in the subsequent discussion.

\subsubsection{QI effects}
First, we want to look at the contributions to the total energy originating from the so-called zero-point vibrations, i.e, vibrations of the ions around their equilibrium positions due to the quantum mechanical treatment of the nuclei, absent in the classical, \textit{clamped-nuclei} picture~\cite{BOA}.
In the Born-Oppenheimer approximation, the total energy $E_\text{tot}[\mathbf{R}]$ (at $T=\SI{0}{K}$) for ionic positions $\mathbf{R}$ is given by the sum of the internal electronic energy $E_\text{el}[\mathbf{R}]$ and the ZPE contributions of the nuclei $E_\text{ZP}[\mathbf{R}]$.
In most solids, $E_\text{ZP}$ is much smaller than $E_\text{el}$ and can be safely neglected. However, due to the small mass of H and the resulting high phonon frequencies in hydrides, the ZPE can become substantial and thus cause a modification of the equilibrium crystal structure.

\begin{figure}[t]
	\includegraphics[width=\columnwidth]{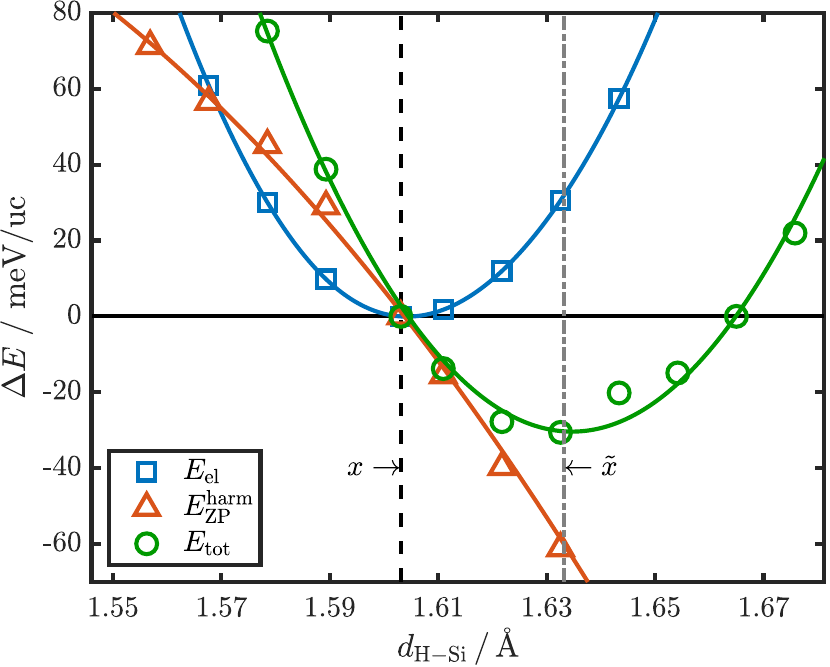}
	\caption{\textbf{Electronic total energy, harmonic ZPE, and total energy:} Electronic total energy $E_\text{el}$ (blue squares), harmonic ZPE $E^\text{harm}_\text{ZP}$ (red triangles), and resulting total energy $E_\text{tot}$ (green circles) as a function of H-Si distance for $a=\SI{6.242}{\angstrom}$, where the DFT minimum $x$ and the \SSCHA\ minimum $\tilde{x}$ are marked explicitly, the latter coinciding with the minimum position of $E_\text{tot}$. The three energy curves are plotted relative to their respective values at $x$, i.e. $\Delta E = E|_{d_\text{H-Si}} - E|_x$. The solid lines represent a cubic spline for $E_\text{el}$ and second order polynomial fits for the other energies.}
	\label{fig:ZPE_vs_free_dist}
\end{figure}

At the harmonic level, the true ZPE can be approximated via $E_\text{ZP}[\mathbf{R}] \approx E^\text{harm}_\text{ZP}[\mathbf{R}] = \int_0^\infty\mathrm{d}\omega\rho_\mathbf{R}(\omega)\hbar\omega/2$, where $\rho_\mathbf{R}(\omega)$ is the DFPT phonon density of states and $\hbar\omega/2$ the ZPE of a quantum harmonic oscillator. 
At constant volume, the only free parameter in the \fmm\ structure is the Wyckoff parameter $x$, defining the H-Si distance, for which we have plotted $E_\text{tot}$, $E_\text{el}$, and $E^\text{harm}_\text{ZP}$ relative to their respective values at the DFT minimum $x$ in Fig.~\ref{fig:ZPE_vs_free_dist} for a lattice constant of $a=\SI{6.242}{\angstrom}$. The results for other lattice constants, i.e., pressures, are provided in Supplementary Fig.~9. We also want to note at this point that for structures not in the DFT minimum, non-vanishing forces occur, at odds with the underlying harmonic approximation. In all the cases considered here, however, the individual atomic force components are still small enough within our computational setup to result in purely real frequencies and sufficiently accurate estimates for the ZPE (see Supplementary Figs.~11 and 12 for further details).

As can be appreciated in this figure, the inclusion of the ZPE, even at the harmonic level, shifts the position of the minimum of the total energy considerably and puts it almost exactly at the minimum position $\tilde{x}$ predicted by the \SSCHA. The differences in $d_\text{H-Si}$ between the \SSCHA\ calculations and the ZPE analysis are, in fact, of the order of \SI{1}{m\angstrom}, i.e., well within the observed stochastic noise in \SSCHA. We want to note that the same is true for \LBH\ (see Supplementary Fig.~10).
Furthermore, inclusion of $E^\text{harm}_\text{ZP}[\mathbf{R}]$ reduces the total energy at its minimum by $\sim$\SI{30}{meV/uc} compared to its value at the DFT minimum, agreeing very nicely with the result from \SSCHA\ ($\sim$\SI{27}{meV/uc}). 

This demonstrates the importance of QI effects of the light H ions on the dynamic stability of the hydride materials, and shows that the minimum structure from \SSCHA\ can already be obtained at the level of harmonic ZPE corrections, at least for this class of materials.

\begin{figure*}[t!]
    \centering
    \includegraphics[width=\textwidth]{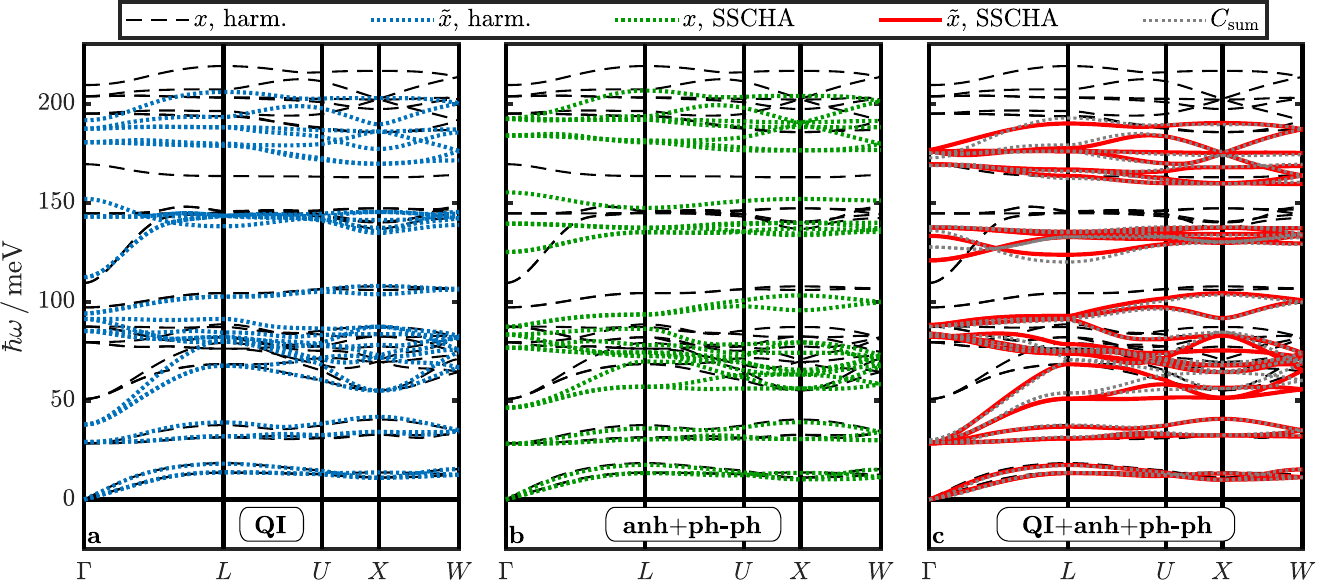}
    \caption{\textbf{Different effects contributing to shifts in the phonon dispersions:} \textbf{a} Harmonic dispersion for atomic positions defined by $x$ (dashed black) and $\tilde{x}$ (dotted blue). \textbf{b} \SSCHA-obtained dispersions for fixed atomic positions defined by $x$ (dotted green lines) and harmonic reference (dashed black). \textbf{c} Dispersions for \SSCHA-relaxed atomic positions defined by $\tilde{x}$ (solid red), obtained from $C_\text{sum}$ (dotted grey, see text for more details), and for the harmonic case (dashed black).}
    \label{fig:effects_ph}
\end{figure*}

Having established that the ZPE has a crucial effect on the structure, we investigate the effect of the changed structure on the phonon dispersions. In Fig.~\mbox{\ref{fig:effects_ph}\textbf{a}}, we present the harmonic dispersions for atomic positions defined by $x$ and $\tilde{x}$. We observe large differences for the high-energy optical modes above \SI{150}{meV}, but also for the low $T_{2g}$ mode at $\Gamma$. The energy shifts for these modes are between 15 and \SI{25}{meV}.

\subsubsection{Anharmonicity and ph-ph interaction effects} Having established the QI effects on the structure and the phonon dispersions, we now want to assess the contributions of phonon anharmonicity (anh) and ph-ph interactions. To do that, we perform a \SSCHA\ calculation while keeping the ions fixed at the DFT equilibrium positions, thus qualitatively removing structural effects on the phonon frequencies.
In that case, we obtain non-vanishing forces within \SSCHA\ (cf. Supplementary Method~2). As the individual force components on the H atoms are still small ($\sim$\SI{150}{meV\per\angstrom} in the cell with $a=\SI{6.242}{\angstrom}$, for example), a fixed-ion calculation seems applicable for a qualitative insight. 
The phonon dispersions obtained from this calculation are presented in Fig.~\mbox{\ref{fig:effects_ph}\textbf{b}}, where we find that all H-dominated optical modes in the whole BZ experience a sizable frequency renormalization. 
It is worth noting that fixed-ions \SSCHA\ calculations indicate the onset of dynamical instability just between 5 and \SI{10}{GPa} (8.4 and \SI{12.2}{GPa}, respectively, for the \SSCHA\ pressure $\tilde{p}$, see Supplementary Fig.~13), which is reasonably close to the frozen-phonon anharmonic (harmonic) \mbox{\pdyn\ = \SI{3(5)}{GPa}}~\cite{Lucrezi2022}.

\subsubsection{Combining QI, anh, and ph-ph effects:} 
As we made the attempt of separating the different contributions from QI effects, anharmonicity, and ph-ph interactions, it is tempting to combine the individual contributions to the phonon dispersion as a simple sum and compare it to the full \SSCHA\ calculation. We approach this task in real space via adding the force constants $C_\text{sum} = C_\text{harm} + \Delta C_\text{QI} + \Delta C_\text{anh/phph}$, where $C_\text{harm}$ is the force constant matrix for harmonic phonons of structure $x$, $\Delta C_\text{QI}$ are the force constant contributions due to the structural changes based on including the ZPE (QI effects), and $\Delta C_\text{anh/phph}$ are the force constant contributions to the anh and ph-ph interaction effects (see Supplementary Note~3).

The dispersions obtained from $C_\text{sum}$ are shown in Fig.~\mbox{\ref{fig:effects_ph}\textbf{c}}, where we also present as reference the phonon dispersions from a full \SSCHA\ calculation.
Again, we find very good agreement between these results, providing further, a posteriori justification and support for the qualitatively introduced separation Ansatz. 
In Table~\ref{tab:effects}, we summarize the considered effects, methods, and the corresponding physical description of the ions.

\begin{table}[h]
	\caption{\textbf{Overview of ionic treatment:} The separate cases are classified according to the structural and phonon treatment. The phonons are obtained either via DFPT or \SSCHA. The ground-state (GS) structure is determined by minimizing either the electronic energy within DFT or the total energy including the ZPE (using DFPT or \SSCHA.)}
	\label{tab:effects}
\begin{tabular}{c||c|c}
$^{\text{\hspace{0.1\columnwidth}phonon}\rightarrow}_{\downarrow\text{structure}}$ & DFPT & \SSCHA \\
 \toprule
$x$ $\hat{=}$ min($E_\text{el}$) & Classical ions & Interact. quantum ions \\
 DFT     &   in el. GS  & in el. GS     \\
         & (standard)   &       (anh+ph-ph)  \\
\hline
 $\tilde{x}$ $\hat{=}$ min($E_\text{tot}$) & Quantum ions  & Interact. quantum ions \\
ZPE/\SSCHA  & in lattice GS  & in lattice GS \\
            &   (QI)         &   (QI+anh+ph-ph)
\end{tabular}
\end{table}

\subsection{Superconductivity}
As \BSH\ is potentially a very promising high-\Tc\ superconductor, we also want to assess the implications of the above mentioned effects on its superconducting (SC) properties.
To do that, we solved the anisotropic Migdal-Eliashberg (ME) equations as implemented in EPW~\cite{ponce_epw_2016} for the four cases in Tab.~\ref{tab:effects}. Details about the calculation within EPW are provided in the Method section and in Supplementary Method~3, at this point we only want to highlight that for each case we used the corresponding force constants to compute the dynamical matrices, and computed the electron-phonon ($ep$) coupling matrix elements as the self-consistent first-order variation of the potential using the equilibrium positions as defined in Tab. \ref{tab:effects}.
In Table~\ref{tab:SCprops}, we summarize the obtained values for quantities characterizing the SC state, i.e., the $ep$ coupling strength $\lambda$, the logarithmic average of the phonon frequencies $\omega_\text{log}$, and the SC critical temperature \Tc.

\begin{table}[h]
	\caption{\textbf{SC properties from ME theory:} Critical temperature \Tc, $ep$ coupling strength $\lambda$, and logarithmic phonon frequency average $\omega_\text{log}$ for the cases discussed in the text.}
	\label{tab:SCprops}
\begin{tabular}{cc|ccc}
effect & (struct., phon.) & $\omega_\mathrm{log}\,/\,\mathrm{meV}$ & $\lambda$ & \Tc\,/\,K\\
 \toprule
standard     & ($x$, harm.)          & 54 & 1.25 & 84  \\
QI           & ($\tilde{x}$, harm.)  & 47 & 1.43 & 82  \\ 
anh+ph-ph    & ($x$, \SSCHA)         & 54 & 1.38 & 90  \\
QI+anh+ph-ph & ($\tilde{x}$, \SSCHA) & 28 & 2.12 & 94  
\end{tabular}
\end{table}

The corresponding Eliashberg spectral functions $\alpha^2F(\omega)$ and the cumulative coupling strengths $\lambda(\omega)$ are shown in Supplementary Fig.~14. We want to stress that the provided values for \Tc \ are obtained by the solution of the full ME equations.
The distribution of the SC gap function $\Delta_\kv$ indicates no change in the distinct two-gap shape calculated for the pure harmonic case~\cite{Lucrezi2022}.
The differences in $\omega_\text{log}$, $\lambda$, and \Tc\ are in the order of 10-15\% except for the full \SSCHA\ calculation, where we see a considerable increase in $\lambda$ to almost double the harmonic value, but also a decrease in $\omega_\text{log}$, compensating the enhancement of $\lambda$. The resulting \Tc\ is increased from 84 to \SI{94}{K}, showing that the full inclusion of all discussed effects results only in $\sim$10-15\% change in \Tc\ for \BSH.

\section{Discussion}
In this work, we study the effects of quantum lattice dynamics within the \SSCHA\ framework on the structure and the dynamical stability of the \fmm\ phase of \BSH. The \SSCHA\ structure relaxation suggests a 2\% elongation of the H-H and H-Si bonds for the studied pressure range of 10 to \SI{30}{GPa} ($\sim\SI{30}{\milli\angstrom}$).

In the phonon dispersions, we find an overall softening of the high optical modes, as well as a dynamic lattice instability characterized by imaginary \SSCHA\ phonon frequencies in the $T_{2g}$ mode at $\Gamma$ below \SI{20}{GPa}, setting the estimate for the critical dynamical pressure to $\tilde{p}_\text{dyn}\approx\SI{20}{GPa}$.
We have further demonstrated the importance of QI effects over anharmonicity and ph-ph interactions, and found that the change in structure, and consequently in pressure, can already be understood by considering harmonic ZPE corrections to the total electronic energy of the system alone (which can be obtained much faster than performing a full \SSCHA\ calculation, cf. Supplementary Note~4). We want to stress at this point that while the total energy ground-state can be obtained already within the harmonic theory, quantum ionic effects and anharmonicity have to be taken into account to accurately determine the corresponding dynamical critical pressure.

We are now left with the question: what is the stability boundary of \fmm-\BSH?
In our previous work on \BSH~\cite{Lucrezi2022}, we challenged the common practice of assuming the range of metastability of high-pressure hydride phases to coincide with the range of (an)harmonic dynamical stability, which systematically underestimates the stabilization pressures needed to synthesize these materials in reality~\cite{PhysRevB.99.220502,Kong2021,Errea2016,Einaga2016}. Dynamical stability is only a prerequisite for thermodynamic metastability, which is characterized by the existence of a distinctive enthalpy barrier that protects a metastable phase from decomposition into other phases (kinetic stability). In our previous work~\cite{Lucrezi2022}, we calculated the enthalpy transition path to the thermodynamic groundstate at different pressures (corresponding to a decomposition of the \fmm\ \BSH\ phase into \mbox{BaSiH$_6$ + H$_2$} in molecular form), and could estimate the barrier height from the intermediate structures (cf. Supplementary Fig.~16, where we also report the minor influence of ZPE on the determined barrier heights). In combination with the calculated convex hulls for the B-S-H system, we can argue with confidence that the \fmm\ \BSH\ phase could be synthesized above \SI{100}{GPa}, and retained down to $\sim\SI{30}{GPa}$, where a distinctive enthalpy barrier still exists. At lower pressures, metastable \fmm\ \BSH\ will decompose, even though (anharmonic) lattice dynamics calculations predict it to be stable. Hence, kinetic stability poses a stricter bound for synthesizability than dynamical stability.

In conclusion, employing ab-initio machine-learned MTPs, we were able to perform \SSCHA\ calculations for \BSH\ at various pressures for supercells up to $4 \times 4 \times 4$ and more than \SI{100000}{} individuals. The inclusion of QI effects, anharmonicity, and ph-ph interactions within the \SSCHA-framework increases the pressure of dynamical stability from \mbox{\pdyn\ $\approx$ \SI{3}{GPa}} to $\tilde{p}_\text{dyn}\approx\SI{20}{GPa}$. We identified the change in structure due to QI effects to be the main driving force here, something that can already be captured to good approximation at the level of harmonic zero-point energy corrections. 

Most importantly, the determined $\tilde{p}_\text{dyn}\approx\SI{20}{GPa}$ is still below \mbox{\pkin\ $\approx$ \SI{30}{GPa}} posed by the concept of kinetic stability, thus the latter represents a much stricter bound for the stability and realizability in these materials.

\section{Methods}
\subsection{DF(P)T calculations}
All DFT and DFPT calculations of electronic and vibrational properties were carried out using the plane-wave pseudopotential code \textsc{Quantum Espresso}~\cite{giannozzi_advanced_2017}, scalar-relativistic optimized norm-conserving Vanderbilt pseudopotentials~\cite{hamann_optimized_2013}, and the \textsc{Pbe}-\textsc{Gga} exchange and correlation functional~\cite{perdew_generalized_1996}. The unit cell calculations are done in the face-centered cubic primitive unit cell with 10 atoms, a 12$\times$12$\times$12 $\kv$-grid, and a plane-wave cutoff energy of \SI{80}{Ry}. The 2$\times$2$\times$2 supercell calculations were done on a 6$\times$6$\times$6 $\kv$-grid. Further details are provided in Supplementary Method~1.

\subsection{\SSCHA\ calculations}
The calculations in the \SSCHA\ are done in the constant-volume relaxation mode, i.e. minimizing the free energy with respect to the average atomic positions $\mathcal{R}$ and the force constants $\Phi$, as implemented in the \SSCHA\ python package \cite{Monacelli_2021}. We use the DFT equilibrium atomic positions and the DFPT dynamical matrices on a 2$\times$2$\times$2 $\qv$-grid as initial guesses for $\mathcal{R}$ and $\Phi$, respectively. The starting point for the larger supercells is obtained by interpolating the previously converged auxiliary dynamical matrices. 

The total energies, forces, and stress tensors for the individuals are obtained from DFT calculations or from machine-learned interatomic potentials in the framework of MTPs, see below. At the end of a minimization run, a new population with higher number of individuals $N$ is generated from the minimized trial density matrix until convergence. We set two stopping criteria for the minimization loops: a Kong-Liu ratio for the effective sample size of 0.2, and ratio of $<10^{-7}$ between the free energy gradient with respect to the auxiliary dynamical matrix and its stochastic error. In calculations based on DFT we increased $N$ up to $10^4$ individuals, for the MTP cases up to $10^5$.

The anharmonic phonon dispersions are obtained from the positional free-energy Hessians without the fourth-order term, if not specified otherwise explicitly. The final atomic positions are obtained from the converged average atomic positions $\mathcal{R}$ and the pressure as derivative of the converged free energy with respect to a strain tensor. 
The free energy difference between the last two populations in the 2$\times$2$\times$2 cells is well below \SI{1}{meV/u.c.}, and well below \SI{0.1}{meV/u.c.} for higher cells. The total forces in the last population are well below $10^{-6}\,\mathrm{meV}\text{\AA}^{-1}$. The physical phonon frequency differences between the last two populations are below \SI{5}{meV} for the DFT cases and well below \SI{1}{meV} for the MTP case ($T_{2g}$ and $A_{2u}$ converge slower, see Supplementary Fig.~6). 
All calculations were carried out at zero temperature.
We note in passing that with these settings, we could reproduce all \LBH\ results from Belli et al.'s work~\cite{Errea_LaBH8}.

Further details are provided in Supplementary Method~2 and details on convergence of the free energy, its gradients, and the auxiliary frequencies in a \SSCHA\ calculation are shown in Supplementary Fig.~15.

\subsection{Moment tensor potentials}
The MTPs were trained and evaluated using the MLIP package \cite{Shapeev2015-MTP,Novikov_2021}. We choose a functional form of level 26, eight radial basis functions, $R_\text{cut}=\SI{5.0}{\angstrom}$ and $R_\text{min}=\SI{1.2}{\angstrom}$, and trained on 50 structures in a 2$\times$2$\times$2 supercell randomly chosen out of all individuals of the DFT \SSCHA\ calculations. We trained separate MTPs for each set of pressure to ensure the highest possible accuracy of our calculations.
We validated the potentials on all individuals in the DFT \SSCHA\ calculations and find a RMSE on the total energy of 0.5-\SI{0.6}{meV/atom}, 45-\SI{50}{\milli\electronvolt\per\angstrom} for the force components, and 0.3-\SI{0.4}{GPa} for the diagonal stress tensor components. We further validated the MTPs on 30 randomly chosen individuals in a 3$\times$3$\times$3 supercell and achieve similar RMSEs. The validations and RMSEs for each pressure are shown in Supplementary Fig.~2 and Supplementary Note~2.

\subsection{ZPE and total energy}
The internal electronic energy $E_\mathrm{el}[\mathbf{R}]$ is obtained from DFT calculations at fixed volume by varying the H-Si distance via the Wyckoff parameter $x$ of the H positions. The phonon density of states $\rho_{\mathbf{R}}(\omega)$ is obtained using DFPT on a 2$\times$2$\times$2 $\qv$-grid, interpolated on a 16$\times$16$\times$16 $\qv$-grid. Smooth ZPE and total energy curves are obtained by second-order polynomial fits in the H-Si distance.
Due to the shift out of the DFT equilibrium structure, forces on the individual H atoms arise at the DFT level. Around the total energy minimum, the force components are in the order of \SI{150}{\milli\electronvolt\per\angstrom}, i.e., small enough to warrant the use of linear-response theory to gain qualitative and systematic insights. DFT diagonal stress tensor components (pressures) are decreased by about \SI{2}{GPa} in the shifted structures around the total energy minimum. The small non-zero forces do not lead to imaginary phonon frequencies in any DFPT calculation for \BSH\ with varied H-Si distance.

Note on similarities to the quasi-harmonic approximation (QHA): Within the QHA, the temperature-dependent internal vibrational energy and the vibrational entropy are commonly added to a system to study temperature effects. The standard QHA only varies external coordinates, such as the volume, but keeps atomic positions in the DFT minimum. In our approach to QI effects, we vary explicitly the internal coordinates at fixed volume, instead. Incidentally, the full optimization of the quasi-harmonic free energy with respect to all degrees of freedom within QHA has just been reported~\cite{masuki2023full}.

\subsection{Migdal-Eliashberg theory}
The Wannier interpolation of the $ep$ matrix elements onto dense $\kv$- and $\qv$-grids, and the subsequent self-consistent solution of the fully anisotropic Migdal-Eliashberg equations were done in \textsc{Epw}~\cite{margine_anisotropic_2013,ponce_epw_2016}, for all the cases in Tab.~\ref{tab:SCprops}. We used coarse 6$\times$6$\times$6 and fine 30$\times$30$\times$30 $\kv$- and $\qv$ grids, a Matsubara frequency cutoff of \SI{1}{eV}, and a Morel-Anderson pseudopotential $\mu^*=0.10$. 
The $ep$ coupling strength
$\lambda = 2\int_0^\infty\mathrm{d}\omega\frac{\alpha^2F(\omega)}{\omega}$ and the logarithmic average phonon frequency $\omega_\text{log} = \exp\left(\frac{2}{\lambda}\int_0^\infty\mathrm{d}\omega\frac{\alpha^2F(\omega)\ln\omega}{\omega}\right)$
are obtained from the Eliashberg spectral function $\alpha^2 F(\omega)$.
Further details are provided in Supplementary Method~3.

\section*{Acknowledgments}
We thank Pedro Pires Ferreira, Carla Verdi, and Luigi Ranalli for fruitful discussions. 
This work was supported by the Austrian Science Fund (FWF), projects P30269 and P32144. Calculations have been performed at the dCluster of TU Graz, as well as on the Vienna Scientific Cluster (VSC4 and VSC5). L.B. acknowledges support from Fondo Ateneo-Sapienza 2018-2022. S.D.C. acknowledges computational resources from CINECA, proj. IsC90-HTS-TECH\_C and IsC99-ACME-C.

\section*{Author contributions}
R.L., E.K., and S.D.C. performed the calculations, M.A. introduced the idea of MTP, and C.H. and L.B. conceived and supervised the project. All authors contributed to the discussion of the results and participated in preparing the manuscript.

\section*{Competing interests}
The authors declare no competing interests.

\section*{Data availability}
The authors confirm that the data supporting the findings of this study are available within the article and its supplementary materials. Further information is available upon request.

\end{document}